\documentclass{article}
\usepackage{fullpage}
\usepackage{graphicx}
\usepackage{amsmath}
\usepackage{amssymb}
\title{Comparison of NNLO DIS scheme splitting functions with results from exact gluon kinematics at small $x$}
\date{}
\renewcommand{\vec}[1]{\mbox{\boldmath$ #1 $}}
\begin{document}
\bibliographystyle{utphys}
\newcommand{\msbar}{\ensuremath{\overline{\text{MS}}}}
\newcommand{\DIS}{\ensuremath{\text{DIS}}}
\setlength{\parindent}{0pt}
\author{C.D. White\thanks{cdw24@hep.phy.cam.ac.uk}, R.S. Thorne\\ \\ Cavendish Laboratory, University of Cambridge,\\ Madingley Road, Cambridge, CB3 0HE, UK}

\maketitle
\thispagestyle{empty}
\begin{abstract}
We consider the effect of exact gluon kinematics in the virtual photon-gluon impact factor at small $x$. By comparing with fixed order DIS scheme splitting and coefficient functions, we show that the exact kinematics results match the fixed order results well at each order, which suggests that they allow for an accurate NLL analysis of proton structure functions. We also present, available for the first time, $x$-space parameterisations of the NNLO DGLAP splitting functions in the DIS scheme, and also the longitudinal coefficients for neutral current scattering.
\end{abstract}

The study of the proton structure functions at small $x$ is of phenomenological importance, given the partonic centre of mass energies now accessible in collider experiments at HERA \cite{H1,ZEUS} and the forthcoming LHC. However, the coefficient and splitting functions relating the structure functions to the parton distributions contain logarithms in the Bjorken $x$ variable. Although QCD fits at next-to-leading order (NLO) in $\alpha_S$ describe the data well, there is some evidence that a resummation of $\log{1/x}$ terms would improve the fits \cite{MRST_errors}. This is accomplished in principle via the BFKL equation \cite{BFKL}, an integral equation for the unintegrated gluon 4-point function $f(x,k_1^2,k_2^2)$ whose kernel is known at next-to-leading logarithmic (NLL) order \cite{BFKL_NLL}. In deep inelastic scattering, the moments of the structure functions ${\cal F}_i(N,Q^2)=\int_0^1x^{N} F_i(x,Q^2)dx$ are then given by the high energy factorisation formula \cite{Collins,CCH}:
\begin{equation}
{\cal F}_i(Q^2,N)=\alpha_S\int_0^\infty\frac{dk^2}{k^2}h_i(k^2/Q^2)f(N,k^2,Q_0^2)g_B(N,Q_0^2),
\label{factorisation}
\end{equation}
where $g_B$ is the bare gluon distribution at momentum scale $Q_0^2$, and the strong coupling $\alpha_S$ is fixed at LL order. The $h_i(k^2/Q^2)$ are the impact factors coupling the virtual photon to the gluon. At present these are known to LL order only \cite{CCH,Catani}. Thus a full NLL order small $x$ analysis of the structure functions is not possible. In section 1 of this paper we show how to obtain the quark-gluon splitting and longitudinal gluon coefficient functions from the impact factors. In section 2, we discuss the DIS scheme coefficient and splitting functions at NNLO, needed to compare directly with the small $x$ expansion, as these have not been presented before. In section 3 we compare the splitting and coefficient functions obtained from (\ref{factorisation}) with the complete NLO and NNLO results, showing that the inclusion of exact gluon kinematics in the photon impact factors gives a good approximation to higher order effects at small $x$, and quite possibly contains the dominant information in a NLL (or higher) order calculation.\\

\section{$P_{qg}^{DIS}$ and $C_{Lg}^{DIS}$ with exact gluon kinematics}
After solving the BFKL equation for $f(N,k^2,Q_0^2)$, the latter two factors in equation (\ref{factorisation}) combine to give a regularised unintegrated gluon distribution, and $h_i$ can naively be interpreted as the coefficient function linking the gluon density with the structure function. However, in the case of $F_2$, the impact factor diverges as $k^2/Q^2\rightarrow 0$. One can understand this given that at ${\cal O}(\alpha_S^0)$, $F_2$ is proportional to the quark singlet parton distribution with no gluon contribution. One does not expect to describe this nonperturbative dependence using perturbation theory, and thus the impact factor diverges. One must instead consider solving the evolution equation for $F_2$, via the quantity:
\begin{equation}
\frac{\partial{\cal F}_2(Q^2,N)}{\partial\ln{Q^2}}=\alpha_S\int_0^\infty\frac{dk^2}{k^2}h_2(k^2/Q^2)f(N,k^2,Q_0^2)g_B(N,Q_0^2),
\label{df2}
\end{equation}
which serves to define the impact factor $h_2$. In a general factorisation scheme, one loses the simple interpretation of $h_2$ as the coefficient function relating the gluon distribution to the structure function. Instead it represents a mixture of the coefficient $C_{2g}$ and the anomalous dimension $\gamma_{qg}$. If one chooses to work in the DIS scheme \cite{Altarelli}, where $F_2$ is given by the naive parton model expression to all orders, then $h_2$ can be interpreted directly as the quark gluon anomalous dimension $\gamma_{qg}^{DIS}$. From equation (\ref{factorisation}), the longitudinal impact factor $h_L$ is identified with the coefficient function $C_{Lg}^{DIS}$ and does not diverge due to the fact the the longitudinal structure function vanishes at ${\cal O}(\alpha_S^0)$. It is convenient to perform a second Mellin transformation on the factorisation formulae to unravel the convolution in $k$-space:
\begin{equation}
\tilde{{\cal F}}_L(\gamma,N)=\int_0^\infty dk^2(k^2)^{-1-\gamma}{\cal F}_L(k^2,N)=\tilde{h}_L(\gamma)\tilde{G}(\gamma,N),
\end{equation}
and similarly for equation (\ref{df2}).\\ 

\begin{figure}
\begin{center}
\scalebox{0.5}{\includegraphics{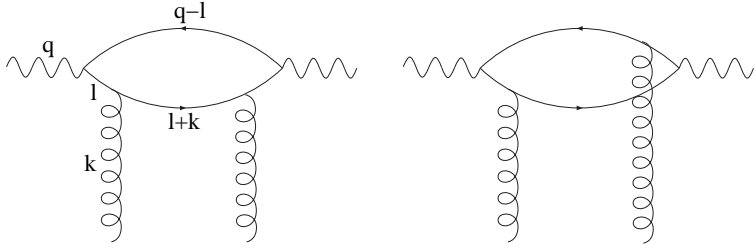}}
\caption{Diagrams contributing to the LL impact factor. Two further diagrams are obtained by reversing the direction of the quark loop.}
\label{impact}
\end{center}
\end{figure}

Diagrams contributing to the impact factor are shown in figure \ref{impact}. 
One may introduce a Sudakov decomposition for the $4$-momenta $k, r$:
\begin{align}
l&=\alpha q' +\beta p +l_\perp;\\
k&=\frac{q^2}{s}q' + x_g p +k_\perp,
\end{align}
where p is the proton $4$-momentum (light-like if one ignores the proton mass), $q'=q+xp$ a second light-like vector involving the Bjorken variable $x=Q^2/(2p\cdot q)$, and $s=(p+q)^2$. The on-shell requirements for the intermediate quarks ($(l+k)^2=(q-l)^2=0$) then lead to the relation:
\begin{equation}
x_g=x\left[\frac{\hat{Q}^2-\hat{k}_\perp^2-{l'_\perp}^2}{\hat{Q}^2}\right],
\label{kinematics}
\end{equation}
with $\hat{A}\equiv\alpha(1-\alpha)A^2$, $Q^2=-q^2$, and $l'_\perp=l_\perp+(1-\alpha)k_\perp$. At LL order, the momentum fraction $x_g$ of the incident proton carried by the gluon is undetermined, as equation (\ref{kinematics}) implies the difference $\log{x_g}-\log{x}$ is finite as $x\rightarrow 0$. By imposing correct kinematics for the gluon, one includes in the impact factor significant higher order information.\\

The resulting $N$ dependent factors $h_2(\gamma,N)$ and $h_L(\gamma,N)$ can be found in \cite{Bialas}, and from them one may derive estimates of $\gamma_{qg}^{DIS}$ and $C_{Lg}^{DIS}$ at fixed order in $\alpha_S$. One first expands the relevant impact factor as a Taylor series in $\gamma$ with coefficients $h_i^{(n)}$. In solving the BFKL equation, $\gamma$ is identified as the anomalous dimension $\gamma(N)$ given at NLL accuracy in \cite{BFKL_NLL} (any further accuracy would require knowledge of the NNLL BFKL kernel). One thus has:
\begin{equation}
C_{Lg}^{DIS(e)}(\alpha_S,N)=\sum_{n=0}^\infty h_L^{(n)}(N)[\gamma(\alpha_S,N)]^{(n)},
\label{clge}
\end{equation}
as the exact kinematics result for the coefficient function up to NLL order, and similarly for $\gamma_{qg}^{DIS(e)}$ in terms of $h_2$. The BFKL anomalous dimension has a perturbative expansion:
\begin{equation}
\gamma(\alpha_S,N)=\sum_{n=1}^{\infty} \alpha_S^n f_n(N),
\label{BFKL_NLL}
\end{equation}
so that equation (\ref{clge}) is a power series in $\alpha_S$, beginning at ${\cal O}(\alpha_S)$. The corresponding $x$-space expressions are given in Appendix A. One may compare order by order with the complete results for $C_{Lg}$ and $\gamma_{qg}$. The corresponding $\overline{\text{MS}}$ functions have been computed up to ${\cal O}(\alpha_S^3)$ \cite{Floratos,Lopez,vanNeerven,Floratos2,Hamberg,Vogt_s,Vogtcl}. However, for a direct comparison with the exact kinematics results one needs the corresponding results in the DIS scheme rather than the conventional $\msbar$ scheme. 

\section{DIS scheme splitting and coefficient functions}
The NNLO singlet and non-singlet splitting functions have recently been computed in the $\overline{\text{MS}}$ scheme \cite{Vogt_s, Vogt_ns}, along with the ${\cal O}(\alpha_S^3)$ coefficient functions for neutral boson exchange \cite{Vogtcl, Vogtc}. They are extremely lengthy - for example, the typeset NNNLO gluon coefficient $C_{2g}$ is around twelve pages long \cite{Vogtc}. The NNLO splitting functions are shorter, but like the coefficients are made more complicated by the nature of the algebraic functions involved. All of the NNLO coefficient and splitting functions involve combinations of harmonic sums in $N$-space, which after inverse Mellin transformation yield harmonic polylogarithms \cite{polylogs} in $x$-space (up to weight five at this order). These are non-standard functions and thus must be generated numerically \cite{numlogs}. The combination of length and numerical complexity makes it is infeasible that the complete results can be immediately used in phenomenological applications. Instead one may parameterise the results in $x$-space in terms of simple algebraic functions, with a precision that far exceeds that due to higher order corrections. The Mellin transforms of the parameterisations then give suitably accurate $N$-space representations. Parameterisations of the $\msbar$ functions are given in \cite{Vogt_s,Vogt_ns, Vogtcl, Vogtc}. From these we have derived corresponding representations of the DIS scheme quantities, accurate to within a percent apart from near the zeros. Our results are presented in Appendix B.\\ 

In transforming between the $\msbar$ and DIS schemes, we follow the argument presented in \cite{vanNeerven,Vogts}. The DIS scheme is characterised by the singlet structure function $F_2$ having the same form as the naive parton model to all orders \cite{Altarelli}:
\begin{equation}
F_{2s}(x,Q^2)=\Sigma^{\DIS}\equiv C^{\msbar}_{2q}\otimes\Sigma^{\msbar}+C_{2g}\otimes g^{\msbar},
\end{equation}
where $\Sigma=\sum_i(q_i+\bar{q}_i)$ is the singlet quark density, and the factorisation and renormalisation scales have been chosen as $Q^2$. The factorisation scheme independence of $F_2$ then imposes a transformation between the DIS and $\msbar$ scheme partons. There remains an ambiguity in the definition of the DIS gluon. However, the momentum sum rule fixes:
\begin{equation}
\int_0^1 dx x[\Sigma(x,Q^2)+g(x,Q^2)]=1
\label{sumrule}
\end{equation}
in both schemes. In Mellin space, this becomes:
\begin{equation}
\Sigma^{\DIS}(N)+g^{\DIS}(N)=\Sigma^{\msbar}(N)+g^{\msbar}(N)
\label{sumruleN}
\end{equation}
for $N=1$\footnote{This corresponds to our choice of Mellin variable and that of \cite{Bialas}. The alternative definition $\tilde{f}(N)=\int_0^1 x^{(N-1)}f(x)$ is also in common use, and in that case the second moment is constrained.}. One may remove the ambiguity by extending equation (\ref{sumruleN}) to all $N$, and one obtains:
\begin{equation}
\vec{q}^{\DIS}\equiv \left(\begin{array}{c}\Sigma^{\DIS}\\g^{\DIS}\end{array}\right)
=\left(\begin{array}{cc}C_{2q}^{\msbar}&C_{2g}^{\msbar}\\-C_{2q}^{\msbar}&-C_{2g}^{\msbar}\end{array}\right)
\otimes\left(\begin{array}{c}\Sigma^{\msbar}\\g^{\msbar}\end{array}\right)\equiv \vec{Z}\otimes\vec{q}^{\msbar}.
\label{MS->DIS}
\end{equation}
To obtain the splitting functions, one differentiates equation (\ref{MS->DIS}) with respect to $Q^2$ and rearranges yielding:
\begin{equation}
\vec{P}^{\DIS}=\left(\vec{Z}\otimes\vec{P}^{\msbar}+\beta(\alpha_S)\frac{d\vec{Z}}{d\alpha_S}\right)\otimes\vec{Z}^{-1}.
\label{splits}
\end{equation}
where $\vec{P}=\left(\begin{array}{cc}P_{qq}&P_{qg}\\P_{gq}&P_{gg}\end{array}\right)$. Substituting the perturbative expansions of the $\msbar$ scheme coefficient and splitting functions\footnote{Conventionally, $P_{ij}^{(n)}$ is the coefficient of $a^{n+1}$, where $a=\alpha_S/(4\pi)$; $C_{\{2,L\}i}^{(n)}$ is the coefficient of $a^n$.}, along with the QCD $\beta$ function\footnote{Here $\beta_n$ is the coefficient of $a^{n+2}$.}, one can derive the DIS scheme results order by order in $\alpha_S$. The explicit transformations at ${\cal O}(\alpha_S^3)$ are:
\begin{align}
P_{qq}^{(2)\DIS}&=P_{qq}^{(2)\msbar}+C_{2q}^{(2)\msbar}\otimes P_{qg}^{(0)\msbar}+C_{2g}^{(2)\msbar}\otimes P_{gq}^{(0)\msbar}+C_{2g}^{(1)\msbar}\otimes P_{gq}^{(1)\msbar}+C_{2q}^{(1)\msbar}\otimes P_{qg}^{(1)\msbar}\notag\\
&-C_{2g}^{(1)\msbar}\otimes P_{gq}^{(0)\msbar}\otimes C_{2q}^{(1)\msbar}+C_{2g}^{(1)\msbar}\otimes P_{gg}^{(0)\msbar}\otimes C_{2q}^{(1)\msbar} -C_{2g}^{(1)\msbar}\otimes P_{qq}^{(0)\msbar}\otimes C_{2q}^{(1)\msbar}\notag\\
&+C_{2g}^{(1)\msbar}\otimes P_{qg}^{(0)\msbar}\otimes C_{2q}^{(1)\msbar}+\beta_0 C_{2q}^{(1)\msbar}\otimes C_{2q}^{(1)\msbar}-\beta_0 C_{2g}^{(1)\msbar}\otimes C_{2q}^{(1)\msbar}-2\beta_0 C_{2q}^{(2)\msbar}\notag\\
&-\beta_1 C_{2q}^{(1)\msbar};
\label{transqq}
\end{align}
\begin{align}
P_{gq}^{(2)\DIS}&=P_{gq}^{(2)\msbar}-C_{2q}^{(2)\msbar}\otimes P_{qq}^{(0)\msbar}-C_{2g}^{(2)\msbar}\otimes P_{gq}^{(0)\msbar}-C_{2q}^{(2)\msbar}\otimes P_{gq}^{(0)\msbar} + C_{2q}^{(2)\msbar}\otimes P_{gg}^{(0)\msbar}\notag\\
&-C_{2q}^{(1)\msbar}\otimes P_{qq}^{(1)\msbar}-C_{2g}^{(1)\msbar}\otimes P_{gq}^{(1)\msbar}-C_{2q}^{(1)\msbar}\otimes P_{gq}^{(1)\msbar}+C_{2q}^{(1)\msbar}\otimes P_{gg}^{(1)\msbar}\notag\\
&+C_{2q}^{(1)\msbar}\otimes P_{qq}^{(0)\msbar}\otimes C_{2q}^{(1)\msbar}+C_{2q}^{(1)\msbar}\otimes P_{gq}^{(0)\msbar}\otimes C_{2q}^{(1)\msbar}-C_{2q}^{(1)\msbar}\otimes P_{gg}^{(0)\msbar}\otimes C_{2q}^{(1)\msbar}\notag\\
&-C_{2q}^{(1)\msbar}\otimes P_{qg}^{(1)\msbar}\otimes C_{2q}^{(1)\msbar}-\beta_0 C_{2q}^{(1)\msbar}\otimes C_{2q}^{(1)\msbar}+\beta_0 C_{2g}^{(1)\msbar}\otimes C_{2q}^{(1)\msbar}+2\beta_0 C_{2q}^{(2)\msbar}\notag\\
&+\beta_1 C_{2q}^{(1)\msbar};
\label{transgq}
\end{align}
\begin{align}
P_{gg}^{(2)\DIS}&=P_{gg}^{(2)\msbar}-C_{2q}^{(2)\msbar}\otimes P_{qg}^{(0)\msbar}-C_{2g}^{(2)\msbar}\otimes P_{gq}^{(0)\msbar}-C_{2q}^{(1)\msbar}\otimes P_{qg}^{(1)\msbar}-C_{2g}^{(1)\msbar}\otimes P_{gq}^{(1)\msbar}\notag\\
&+C_{2g}^{(1)\msbar}\otimes P_{qq}^{(0)\msbar}\otimes C_{2q}^{(1)\msbar}+C_{2g}^{(1)\msbar}\otimes P_{gq}^{(0)\msbar}\otimes C_{2q}^{(1)\msbar}-C_{2g}^{(1)\msbar}\otimes P_{gg}^{(0)\msbar}\otimes C_{2q}^{(1)\msbar}\notag\\
&-C_{2g}^{(1)\msbar}\otimes P_{qg}^{(0)\msbar}\otimes C_{2q}^{(1)\msbar}+\beta_0C_{2g}^{(1)\msbar}\otimes C_{2g}^{(1)\msbar}-\beta_0 C_{2q}^{(1)\msbar}\otimes C_{2g}^{(1)\msbar}+2\beta_0 C_{2g}^{(2)\msbar}\notag\\
&+\beta_1 C_{2g}^{(1)\msbar};
\label{transgg}
\end{align}
\begin{align}
P_{qg}^{(2)\DIS}&=P_{qg}^{(2)\msbar}+C_{2q}^{(2)\msbar}\otimes P_{qg}^{(0)\msbar}+C_{2g}^{(2)\msbar}\otimes P_{gg}^{(0)\msbar} -C_{2g}^{(2)\msbar}\otimes P_{qq}^{(0)\msbar}+C_{2g}^{(2)\msbar}\otimes P_{qg}^{(0)\msbar}\notag\\
&+C_{2q}^{(1)\msbar}\otimes P_{qg}^{(1)\msbar}+C_{2g}^{(1)\msbar}\otimes P_{gg}^{(1)\msbar}-C_{2g}^{(1)\msbar}\otimes P_{qq}^{(1)\msbar}+C_{2g}^{(1)\msbar}\otimes P_{qg}^{(1)\msbar}\notag\\
&-C_{2g}^{(1)\msbar}\otimes P_{gq}^{(0)\msbar}\otimes C_{2g}^{(1)\msbar}+C_{2g}^{(1)\msbar}\otimes P_{gg}^{(0)\msbar}\otimes C_{2g}^{(1)\msbar}-C_{2g}^{(1)\msbar}\otimes P_{qq}^{(0)\msbar}\otimes C_{2g}^{(1)\msbar} \notag\\
&+C_{2g}^{(1)\msbar}\otimes P_{qg}^{(0)\msbar}\otimes C_{2g}^{(1)\msbar}+\beta_0 C_{2q}^{(1)\msbar}\otimes C_{2g}^{(1)\msbar}-\beta_0 C_{2g}^{(1)\msbar}\otimes C_{2g}^{(1)\msbar}-2\beta_0 C_{2g}^{(2)\msbar}\notag\\
&-\beta_1 C_{2g}^{(1)\msbar}.
\label{transqg}
\end{align}
Non-singlet quark combinations transform according to:
\begin{equation}
q^{\DIS}_{ns}=C_{2ns}^{\msbar}\otimes q_{ns}^{\msbar},
\label{transqns}
\end{equation}
which has the form of equation (\ref{MS->DIS}) but with a trivial transformation matrix. Hence one obtains for the non-singlet splitting functions relevant to neutral and charged current scattering \cite{Vogt_ns}:\\
\begin{equation}
P_{ns}^{+,-(2)\DIS}=P_{ns}^{+,-(2)\msbar}+\beta_0 C_{2q}^{(1)\msbar}\otimes C_{2q}^{(1)\msbar}-2\beta_0 C_{2ns}^{+,-(2)\msbar}-\beta_1 C_{2q}^{(1)\msbar};
\label{transns}
\end{equation}
The pure singlet splitting function is given by:
\begin{equation}
P_{ps}^{(2)\DIS}=P_{qq}^{(2)\DIS}-P_{ns}^{+(2)\DIS}.
\label{transps}
\end{equation}

The $F_2$ coefficient functions are simply defined to all orders in the DIS scheme. For the longitudinal coefficients, one considers:
\begin{equation}
F_L=\left(\begin{array}{ccc}C^{\msbar}_{Lq}&C^{\msbar}_{Lg}&C^{\msbar}_{Lns}\end{array}\right)\otimes
\left(\begin{array}{c}\Sigma^{\msbar}\\ g^{\msbar}\\ q_{ns}^{\msbar}\end{array}\right).
\end{equation}
Using the transformation equations (\ref{MS->DIS}) and (\ref{transqns}), one finds:
\begin{equation}
\left(\begin{array}{ccc}C^{\DIS}_{Lq}&C^{\DIS}_{Lg}&C^{\DIS}_{Lns}\end{array}\right)
=\left(\begin{array}{ccc}C^{\msbar}_{Lq}&C^{\msbar}_{Lns}&C^{\msbar}_{Lg}\end{array}\right)\otimes 
\left(\begin{array}{cc}\vec{Z}&0\\0&C_{2ns}^{+\msbar}\end{array}\right)^{-1}.
\label{MS->DISc}
\end{equation}
Explicit results at ${\cal O}(\alpha_S^3)$ after substituting the expansions of the $\msbar$ scheme coefficient functions are:
\begin{align}
C_{Lg}^{(3)\DIS}&=C_{Lg}^{(3)\msbar}+C_{Lg}^{(1)\msbar}\otimes C_{2g}^{(2)\msbar}-C_{Lq}^{(1)\msbar}\otimes C_{2g}^{(2)\msbar}+C_{Lg}^{(2)\msbar}\otimes C_{2g}^{(1)\msbar}-C_{Lq}^{(2)\msbar}\otimes C_{2g}^{(1)\msbar}\notag\\
&+C_{Lg}^{(1)\msbar}\otimes C_{2g}^{(1)\msbar}\otimes C_{2g}^{(1)\msbar}-C_{Lq}^{(1)\msbar}\otimes C_{2g}^{(1)\msbar}\otimes C_{2g}^{(1)\msbar}-C_{Lg}^{(1)\msbar}\otimes C_{2q}^{(1)\msbar}\otimes C_{2g}^{(1)\msbar}\notag\\
& +C_{Lq}^{(1)\msbar}\otimes C_{2q}^{(1)\msbar}\otimes C_{2g}^{(1)\msbar};
\label{transclg}
\end{align}
\begin{align}
C_{Lq}^{(3)\DIS}&=C_{Lq}^{(3)\msbar}+C_{Lg}^{(1)\msbar}\otimes C_{2q}^{(2)\msbar}-C_{Lq}^{(1)\msbar}\otimes C_{2q}^{(2)\msbar}+C_{Lg}^{(2)\msbar}\otimes C_{2q}^{(1)\msbar}-C_{Lq}^{(2)\msbar}\otimes C_{2q}^{(1)\msbar}\notag\\
&+C_{Lg}^{(1)\msbar}\otimes C_{2g}^{(1)\msbar}\otimes C_{2q}^{(1)\msbar}-C_{Lq}^{(1)\msbar}\otimes C_{2g}^{(1)\msbar}\otimes C_{2q}^{(1)\msbar}-C_{Lg}^{(1)\msbar}\otimes C_{2q}^{(1)\msbar}\otimes C_{2q}^{(1)\msbar}\notag\\
&+C_{Lq}^{(1)\msbar}\otimes C_{2q}^{(1)\msbar}\otimes C_{2q}^{(1)\msbar};
\label{transclq}
\end{align}
\begin{equation}
C_{Lns}^{+(3)\DIS}=C_{Lns}^{(3)\msbar}-C_{Lns}^{(2)\msbar}\otimes C_{2ns}^{(1)\msbar}-C_{Lns}^{(1)\msbar}\otimes C_{2ns}^{(2)\msbar}+C_{Lns}^{(1)\msbar}\otimes C_{2ns}^{(1)\msbar}\otimes C_{2ns}^{(1)\msbar}.
\label{transclns}
\end{equation}
Then the pure singlet coefficient is given by:
\begin{equation}
C_{Lps}^{(3)\DIS}=C_{Lqq}^{(3)\DIS}-C_{Lns}^{+(3)\DIS}.
\label{transclps}
\end{equation}

The transformation terms were evaluated in $N$-space, and divergent high and low $N$ limits were then extracted. Up to ${\cal O}(1/N)$, one has a choice in how to extract the high $N$ piece. We have chosen this in such a way as to lead to simple plus distributions and logarithms of $(1-x)$ in the $x$-space functions. The remaining finite functions as $N\rightarrow 0,\infty$ were parameterised in $x$-space by evaluating the inverse Mellin transform numerically. Finally the transformation terms were added to the existing $\overline{\text{MS}}$ parameterisations. Thus the plus distribution and small-$x$ divergent terms are exact up to truncation of the coefficients, as also are the parts of the $\log(1-x)$ terms not involving $(1-x)\log(1-x)$.\\

The coefficients of $\delta(1-x)$ in $P_{ns}^+$, $P_{gq}$ and $P_{gg}$ have been modified, and $\delta(1-x)$ contributions added to $P_{qg}$ that should in principle be absent. This, following refs. \cite{Vogt_s,Vogt_ns,Vogtc,Vogtcl}, is to increase the $N$-space accuracy, such that the parameterised functions satisfy the momentum sum rules:
\begin{align}
\gamma_{qg}^{(2)DIS}(N)+\gamma_{gg}^{(2)DIS}(N)&=0;\label{sum1}\\
\gamma_{gq}^{(2)DIS}(N)+\gamma_{qq}^{(2)DIS}(N)&=0,\label{sum2}
\end{align}
for $N=1$ \footnote{This also implies that the $n_f$ independent part of $P_{gg}^{(2)DIS}$ should vanish, given that $P_{qg}$ has no term at ${\cal O}(n_f^0)$.}. One can also introduce such terms into the longitudinal coefficient functions, by fitting to numerical values of the low integer moments. We choose not to introduce these, however, given the size of these effects (no more than a few parts permille) do not exceed the uncertainty of the parameterisations. We have checked all of our expressions against known numerical moments \cite{Retey}.\\
 
Particularly noteworthy is the singularity structure of the DIS scheme functions as $x\rightarrow 1$. One sees that the singlet quark splitting functions contain plus distributions up to ${\cal D}_2$ \footnote{See Appendix B for the definition of these functions.}, or $\log^3(N)$ in Mellin space. However, $P_{gq}^{(2)\DIS}$ contains more singular terms up to ${\cal D}_4\equiv\log^5(N)$. One can understand this by considering what happens in the $\overline{\text{MS}}$ scheme. There $\log(N)$ terms arise in the coefficients $C_{2q}$ and $C_{2ns}$ as a result of soft gluon emission from the quark probed by the virtual photon. As $x\rightarrow 1$, there is insufficient phase space for the emission of real gluons, and thus an incomplete cancellation between singularities arising from virtual and real emission. The leading logarithms in $N$ exponentiate \cite{exp1, exp2}, and the sub-leading logarithms can also be resummed \cite{Sterman,Kidonakis,Catani_largex,Webber}. Combining the known resummation and fixed order results allows knowledge of the four leading towers of high $N$ logarithms in $C_{2q}$ to all orders in $\alpha_S$ \cite{Vogt_soft}\footnote{This analysis has very recently been extended to include even higher order logarithmic corrections to DIS and Drell-Yan type processes \cite{Vogt_largex}.}. In the DIS scheme there are no such logarithms in the coefficients, as $C_{2q}$ is defined trivially to all orders. Instead the soft gluon resummation effects enter the splitting functions. The leading $\log(N)$ terms in $C_{2ns}^{\overline{\text{MS}}}$ are produced by exponentiating those in $\gamma_{ns}^{\DIS}$. This follows from the $N$-space evolution equation for the non-singlet quark density:
\begin{equation}
\frac{\partial\tilde{q}_{ns}^{\DIS}}{\partial\log{Q^2}}=\gamma_{ns}^{\DIS}(N)\tilde{q}_{ns}^{\DIS},
\label{qns}
\end{equation}
which is easily solved to give:
\begin{align}
\tilde{q}_{ns}^{\DIS}(Q^2)&=\tilde{q}_{ns}^{\DIS}(Q_0^2)\exp{\left[\int_{\alpha_S(Q_0^2)}^{\alpha_S(Q^2)}\gamma_{ns}^{\DIS}\frac{d\alpha_S}{\beta(\alpha_S)}\right]}\notag\\
&= q_{ns}^{\DIS}(Q_0^2)\exp{\left[\int_{\alpha_S(Q_0^2)}^{\alpha_S(Q^2)}\sum_{n=0}^\infty\sum_{m=0}^\infty\int_{\alpha_S(Q_0^2)}^{\alpha_S(Q^2)} c_{n,m} \beta_m \gamma_{ns}^{(n)\DIS} \alpha_S^{m+n-1}d\alpha_S\right]},
\label{qns_soln}
\end{align}
where the $c_{n,m}$ are coefficients obtained after substituting in the perturbative expansions of the $\beta$ function and anomalous dimension. Performing the integration in the exponent gives:
\begin{equation}
q_{ns}^{\DIS}(Q^2)=q_{ns}^{\DIS}(Q_0^2)\left[\frac{\alpha_S(Q^2)}{\alpha_S(Q_0^2)}\right]^{-\frac{\gamma_{ns}^{(0)DIS}}{\beta_0}} \exp\left\{-\left(\frac{\gamma_{ns}^{(1)DIS}}{\beta_0}+\frac{\beta_1\gamma_{ns}^{(0)DIS}}{\beta_0^2}\right)\left[\frac{\alpha_S(Q^2)}{4\pi}-\frac{\alpha_S(Q_0^2)}{4\pi}\right]+\ldots\right\},
\label{expon}
\end{equation}
where the ellipsis denotes terms giving rise to sub-leading logarithms. Given that $\gamma_{ns}^{(0)DIS}\sim \log{N}$ and $\gamma_{ns}^{(1)DIS}\sim \log^2(N)$ as $N\rightarrow\infty$, the leading logarithms in the exponent come from the term in $\gamma_{ns}^{(1)}$. The form of the non-singlet structure function in the DIS and $\msbar$ schemes is:
\begin{equation}
F_{2ns}=q_{ns}^{\DIS}\equiv C_{2ns}^{\msbar} \otimes q_{ns}^{\msbar}.
\label{F2ns}
\end{equation}
Thus from equation (\ref{qns_soln}),  ones sees that the leading powers of $\log(N)$ in the $\msbar$ scheme non-singlet coefficient function are generated by exponentiation of those in the DIS scheme NLO anomalous dimension $\gamma^{(1)\DIS}_{ns}$ (the LO anomalous dimension is independent of the factorisation scheme, and thus the prefactor in equation (\ref{expon}) is also found in the $\msbar$ scheme). The next-to-leading $\log(N)$ terms in $C_{ns}^{\msbar}$ are not so straightforward, but are determined by the exponentiation of a mixture of $\gamma_{ns}^{(1)DIS}$ and $\gamma_{ns}^{(2)DIS}$, and so on for the other sub-leading logarithms. A similar argument relates the leading $\log(N)$ terms in $C_{2q}^{\msbar}$ with those in $\gamma_{qq}^{(1)\DIS}$.\\

This explains the absence of more singular logarithms $\sim\log^4(N)$ in the DIS scheme $\gamma_{qq}^{(2)}$, as the highest power of $\log(N)$ is limited by the fact that it cannot exceed the power obtained by exponentiation of the leading log term in $\gamma_{qq}^{(1)\DIS}$. Looking at equation (\ref{transqq}), the transformation terms in $P_{qq}^{(2)\overline{\text{MS}}}\rightarrow P_{qq}^{(2)\DIS}$ involve the combination $\beta_0 C_{2q}^{(1)\msbar}\otimes C_{2q}^{(1)\msbar} - 2\beta_0 C_{2q}^{(2)\msbar}$. Thus $\log^4(N)$ terms in $\tilde{C}_{2q}^{(2)\msbar}$ are cancelled by the combination $[\tilde{C}_{2q}^{(1)\msbar}]^2/2!$ due to the exponential structure of the leading logs in the coefficient function. \\

The ${\cal D}_4$ term in $P_{gq}^{(2)\DIS}$ corresponds to a next to leading high $x$ divergence in $C_{2q}^{\msbar}$ ($\sim \alpha_S^3\log^5(N)$ in Mellin space), arising from the terms in equation (\ref{transgq}):\\
\begin{align}
\left[P_{gq}^{(2)DIS}\right]_{{\cal D}_4}&=\left[-C_{2q}^{(2)\msbar}\otimes P_{qq}^{(0)\msbar} + C_{2q}^{(0)\msbar}\otimes P_{gg}^{(0)\msbar}+C_{2q}^{(1)\msbar}\otimes P_{qq}^{(0)\msbar}\otimes C_{2q}^{(1)\msbar}\right.\notag\\
&\left.-C_{2q}^{(1)\msbar}\otimes P_{gg}^{(0)\msbar}\otimes C_{2q}^{(1)\msbar}\right]_{{\cal D}_4}.
\label{d4}
\end{align}

The small $x$ limit of $P_{qg}^{(2)\DIS}$ will be  discussed in section 3 of this paper. Looking at the other splitting functions, one may verify the LL relations \cite{Catani}:
\begin{eqnarray}
P_{gq}=\frac{C_F}{C_A}P_{gg}, & P_{qq}=\frac{C_F}{C_A}\left[P_{qg}-\frac{\alpha_S}{2\pi}T_R\frac{2}{3}\right],
\end{eqnarray}
where $C_A=3$, $C_F=4/3$ are the QCD Casimir invariants and $T_R=1/2$. These relations are also true at LL order in the $\msbar$ scheme.\\

\begin{figure}
\begin{center}
\scalebox{0.8}{\includegraphics{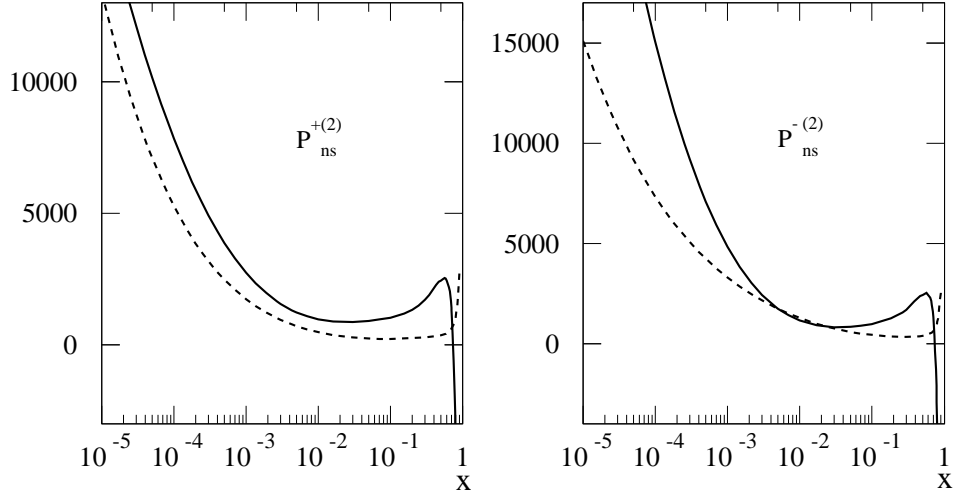}}
\caption{The non-singlet splitting functions in the DIS (solid) and $\msbar$ schemes (dashed), for $n_f=4$.}
\label{non-singlet}
\end{center}
\end{figure}

\begin{figure}
\begin{center}
\scalebox{0.8}{\includegraphics{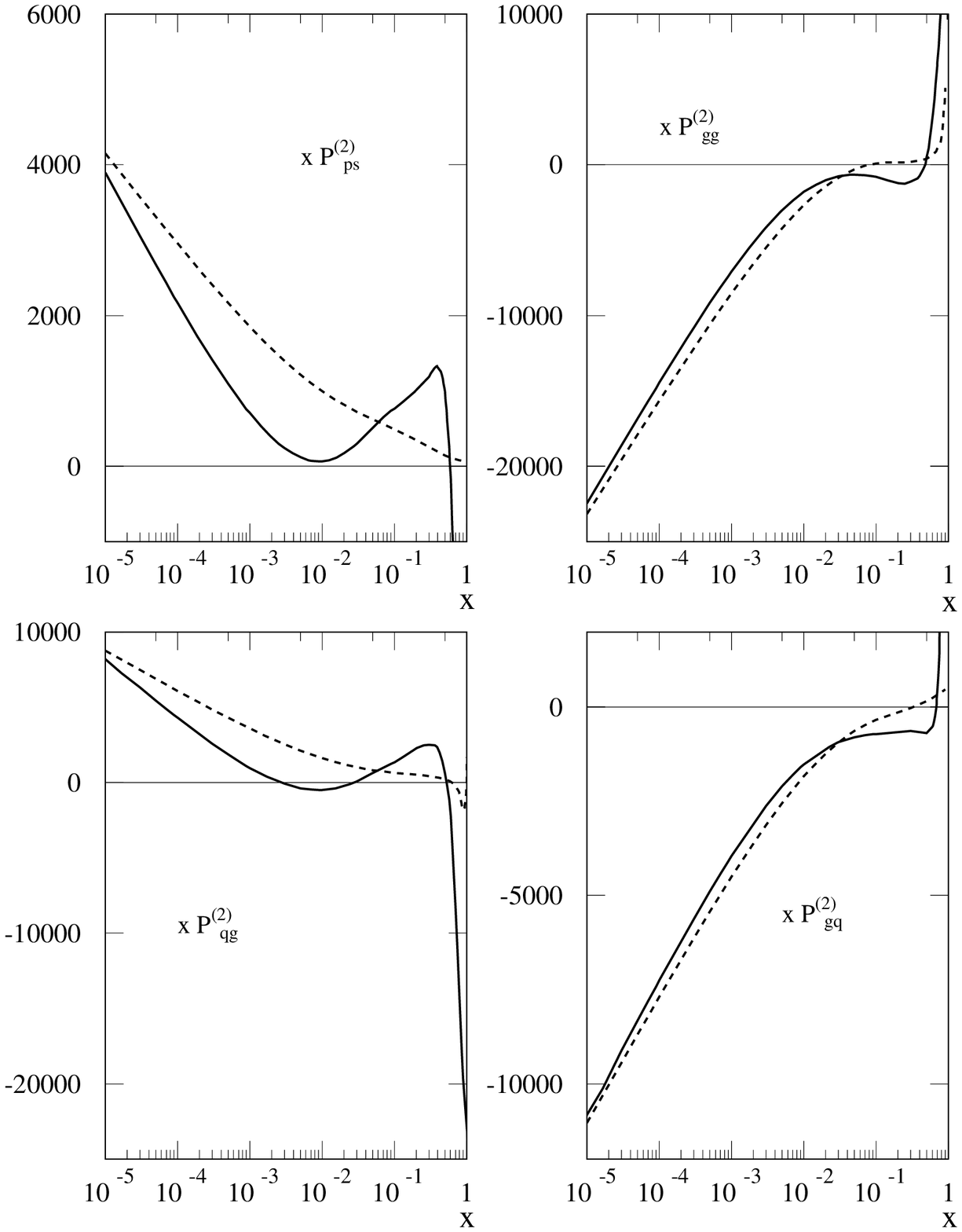}}
\caption{The singlet splitting functions in the DIS (solid) and $\msbar$ schemes (dashed), multiplied by $x$ due to the divergence at small $x$.}
\label{singlet}
\end{center}
\end{figure}

The non-singlet and singlet splitting functions are plotted in figures \ref{non-singlet} and \ref{singlet} respectively. The singlet functions have been multiplied by $x$ to alleviate the small $x$ divergence. 
Aside from the differences at high $x$ discussed above, one sees that the DIS scheme functions are more divergent at small $x$. This is analogous to the high $x$ behaviour -  in changing schemes one transfers divergences from the quark singlet coefficient to the splitting functions. Note also the qualitatively different structures at intermediate $x$ in the two schemes. Each of the singlet splitting functions develops an extra turning point in the DIS scheme.\\

The NNLO $P_{qg}$ develops a negative dip in the DIS scheme at intermediate $x$, before increasing again as $x\rightarrow 0$. Together with the large negative dip at high $x$ this gives a negative result at intermediate $x$ when convolved with a model gluon distribution which is more singular than the splitting function. We return to this feature in section 3. In fact, the qualitative structure of the DIS scheme splitting function can be reproduced from the truncated transformation:
\begin{equation}
P_{qg}^{(2)\DIS}\sim P_{qg}^{(2)\msbar}+C_{2g}^{(2)\msbar}\otimes P_{gg}^{(0)\msbar}-2\beta_0 C_{2g}^{(2)\msbar}.
\label{trunctrans}
\end{equation}

\begin{figure}
\begin{center}
\scalebox{0.8}{\includegraphics{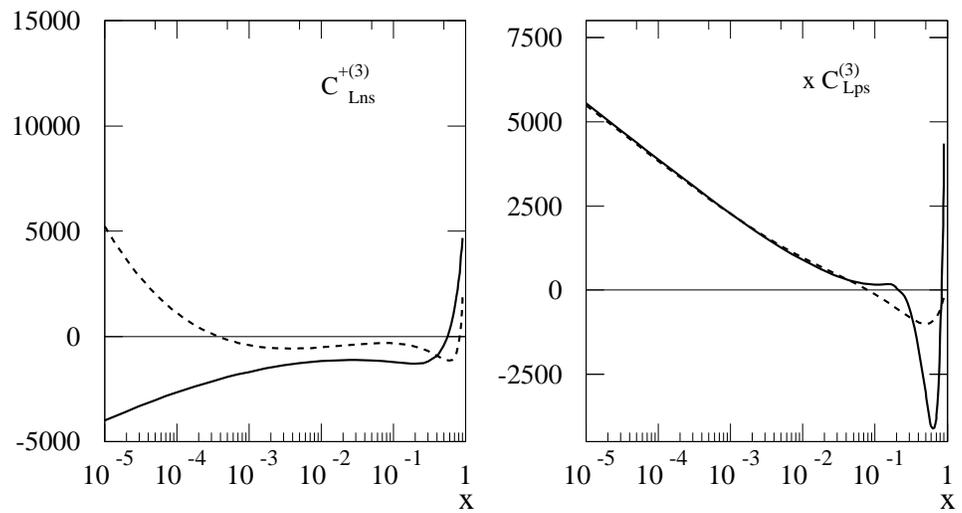}}
\caption{The longitudinal quark coefficient functions in the DIS (solid) and $\msbar$ schemes (dashed), for $n_f=4$. The pure singlet coefficient has been multiplied by $x$.}
\label{quark}
\end{center}
\end{figure}

\begin{figure}
\begin{center}
\scalebox{0.8}{\includegraphics{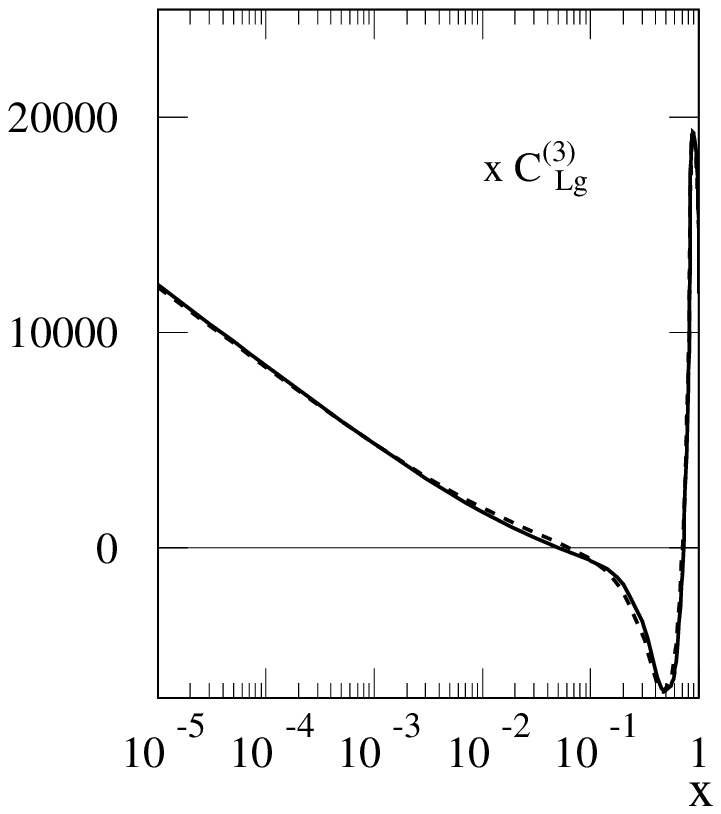}}
\caption{The longitudinal gluon coefficient function in the DIS (solid) and $\msbar$ schemes (dashed) with $n_f=4$, multiplied by $x$ due to the divergence at small $x$.}
\label{gluon}
\end{center}
\end{figure}

\begin{figure}	
\begin{center}
\scalebox{0.8}{\includegraphics{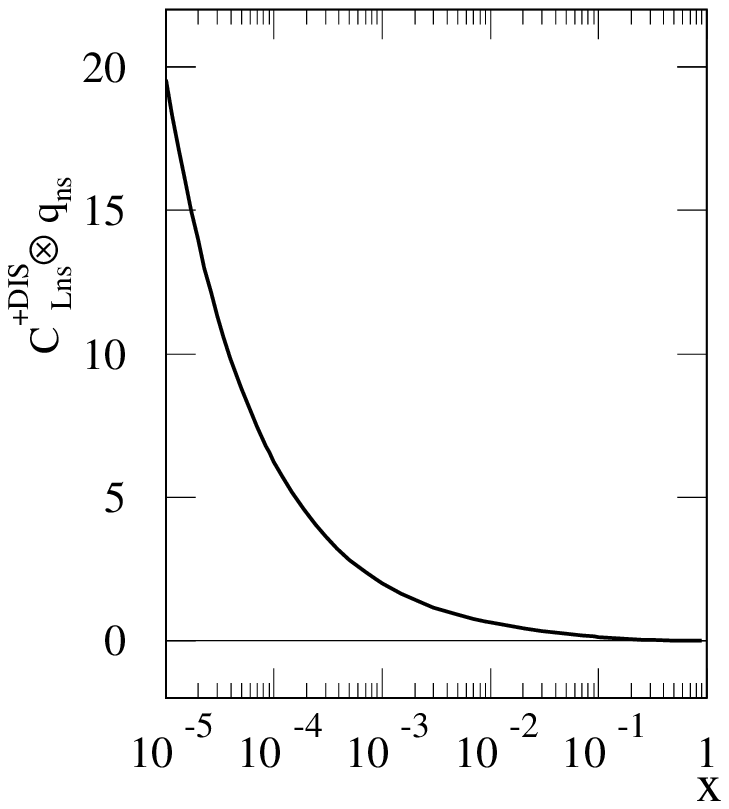}}
\caption{$C_{Lns}^{+\DIS}$ to ${\cal O}(\alpha_S^3)$ convolved with $q_{ns}=x^{-0.5}(1-x)^3$ (note an additional factor of $x$ is not included). Despite the negative sign of the coefficient at low $x$, the structure function $F_{Lns}\sim xC_{Lns}\otimes q_{ns}$ is positive.}
\label{flns}
\end{center}
\end{figure}

The longitudinal quark and gluon coefficient functions are shown in figures \ref{quark} and \ref{gluon}. 
The two gluon coefficients are extremely similar. The pure singlet and gluon coefficients share the same small $x$ limit, as the LL coefficients are the same in both schemes. There is, however, an extra turning point in the DIS scheme pure singlet function at higher $x$. Also of note is the negativity of the non-singlet coefficient at small $x$, a property also shared by the NLO result such that the complete non-singlet coefficient is negative at small $x$. However, it is not divergent as $x\rightarrow 0$ so that convolution with a suitable non-singlet test function does not give a negative non-singlet structure function (see figure \ref{flns}).\\

We now compare the DIS scheme quark-gluon anomalous dimension and longitudinal gluon coefficient with the corresponding results derived from exact gluon kinematics.

\section{Comparison of exact kinematics with NLO and NNLO results}
\begin{figure}
\begin{center}
\scalebox{0.8}{\includegraphics{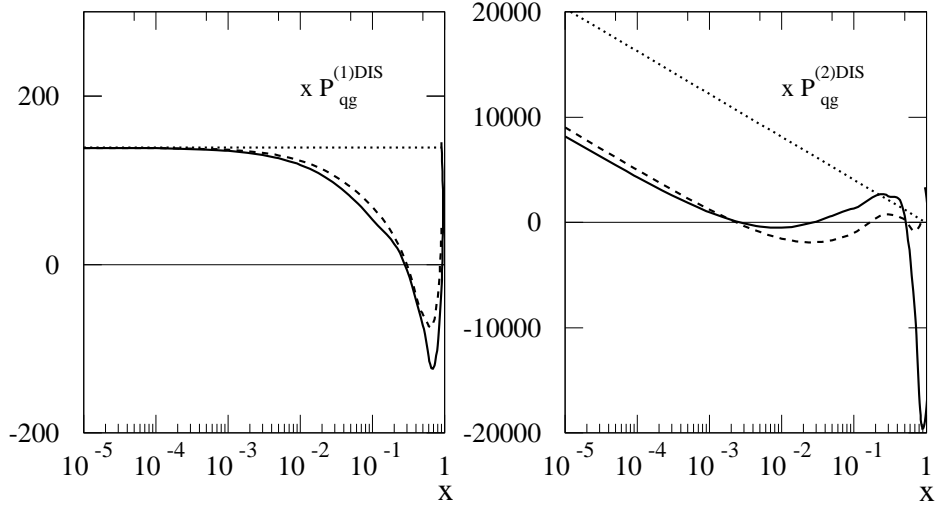}}
\caption{The DIS scheme NLO and NNLO quark-gluon anomalous dimensions (solid) (multiplied by $x$ to remove the small $x$ divergence) alongside the results obtained from the LL photon-gluon impact factor with exact gluon kinematics (dashed). Also shown are the LL approximations (dotted).}
\label{GAMqg}
\end{center}
\end{figure}

At ${\cal O}(\alpha_S)$, the exact kinematics results correspond exactly with the complete results as at this order, all the relevant diagrams are included in the impact factor calculation. The imposition of exact kinematics then supplements the $x$ dependence that is missing when evaluating these diagrams in the LL limit. At higher orders, one can compare the complete $N$-space functions with the estimates obtained from the modified impact factors. Expanding in $N$ one finds:
\begin{align}
\gamma_{qg}^{(1) DIS(e)}&=\frac{34.67n_f}{N}-(102.9n_f+.2140n_f^2)+(172.1n_f+.4391n_f^2)N\notag\\
&-(246.0n_f+.6598n_f^2)N^2+{\cal O}(N^3);\\
\gamma_{qg}^{(2) DIS(e)}&=\frac{441.47n_f}{N^2}-\frac{2635n_f+49.53n_f^2}{N}+(7555n_f+118.0n_f^2+.01682n_f^3)\notag\\
&-(15089n_f+204.0n_f^2+.06958n_f^3)N+(25166n_f+312.1n_f^2\notag\\
&+.1462n_f^3)N^2+{\cal O}(N^3),
\end{align}
The complete NLO and NNLO results give:
\begin{align}
\gamma_{qg}^{(1)DIS}&=\frac{34.67n_f}{N}+(-109.3n_f+.8889n_f^2)+(233.6n_f-5.072n_f^2)N\notag\\
&+(-374.6n_f+11.70n_f^2)N^2+{\cal O}(N^3); \label{expandqg2}\\
\gamma_{qg}^{(2)DIS}&=\frac{441.47n_f}{N^2}+\frac{-3165n_f+30.19n_f^2}{N}+(12945n_f-399.5n_f^2+.5926n_f^3)\notag\\
&+(-34493n_f+1589n_f^2-6.121n_f^3)N\notag\\
&+(73141n_f-4389n_f^2+27.53n_f^3)N^2+{\cal O}(N^3).\label{expandqg3}
\end{align}
The leading logarithms in $x$ (most divergent terms as $N\rightarrow 0$) are correctly predicted from the resummation, and one sees that the next to leading terms in $\gamma_{qg}^{(2)DIS}$ are well estimated by the exact kinematics expression (within 2\% at NLO and 7\% at NNLO, for $n_f=4$). Accuracy falls off for higher order terms in $N$, although these are not associated with small $x$ divergence. The $x$-space functions are shown in figure \ref{GAMqg}. 
The exact kinematics results qualitatively reproduce the structures of the complete results, even at high $x$. They clearly do much better than the LL terms at approximating the splitting functions. Note that the small-$x$ behaviour does not set in until rather low $x$, as can be seen by the splitting function only turning positive for $x\lesssim 2\times 10^{-3}$ at NNLO (for $n_f=4$). The qualitative trend is that at higher order in $\alpha_S$, the splitting function turns positive at lower $x$. We have confirmed, for example,  that the NNNLO exact kinematics splitting function does not turn positive until $x\lesssim 10^{-4}$. A good estimate for these values is obtained by approximating the exact kinematics splitting functions by their asymptotic limits as $x \rightarrow 0$:
\begin{align}
xP_{qg}^{(2)DIS(e)}&\simeq 441.47n_f\log\frac{1}{x}-(2635n_f+49.53n_f^2)+\ldots;\\
xP_{qg}^{(3)DIS(e)}&\simeq 11671n_f\frac{1}{2!}\log^2\frac{1}{x}+(-78095n_f-1410n_f^2)\log\frac{1}{x}+(248414n_f+6924n_f^2+8.265n_f^3)+\ldots,
\end{align}
where the ellipses represent terms vanishing in this limit. Setting $xP_{qg}^{(n)DIS(e)}=0$ gives the approximate value $x=x_0$ at which the LL terms begin to dominate over the sub-leading logarithms. For $n_f=4$, one finds $x_0\simeq 3\times 10^{-3}, 1\times 10^{-4}$ at NNLO, NNNLO respectively. The lower value of $x_0$ with increasing order of $\alpha_S$ implies that the leading small-$x$ resummation effects become less important phenomenologically at higher orders, as sub-leading logarithms dominate until very small $x$.\\

\begin{figure}
\begin{center}
\scalebox{0.5}{\includegraphics{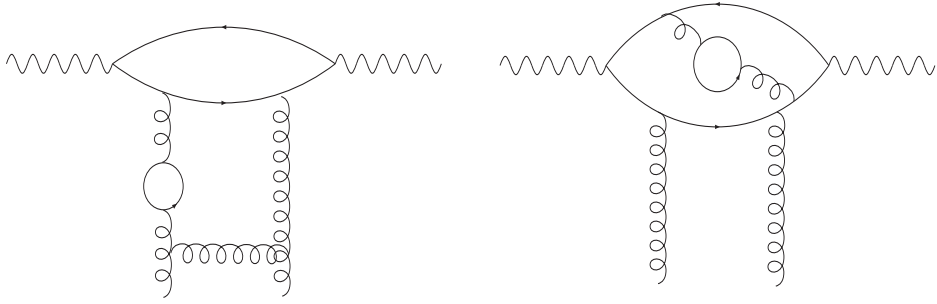}}
\caption{Example diagrams which contribute at ${\cal O}(\alpha_S^3n_f^2)$ to the structure functions.}
\label{bubbles}
\end{center}
\end{figure}

\begin{figure}
\begin{center}
\scalebox{0.5}{\includegraphics{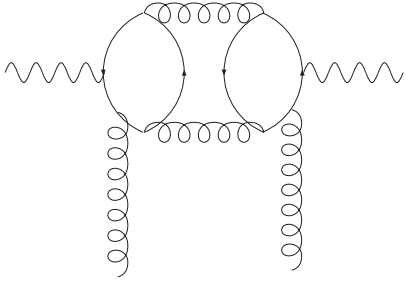}}
\caption{Diagram contributing a term $\propto fl_{11}^g$ to the structure function.}
\label{fl11g}
\end{center}
\end{figure}

From equations (\ref{expandqg2}) and (\ref{expandqg3}), we note that in the non-leading logarithmic terms, contributions involving higher powers of $n_f$ are estimated poorly - including being of the wrong sign. This is expected given that higher powers of $n_f$ in the perturbative contribution to the structure functions may arise from diagrams such as those shown in figure \ref{bubbles}, with fermion bubbles in the vertical rungs of the gluon ladder and in the quark loop at the top of the diagram. The former are included in the NLL BFKL anomalous dimension\footnote{Fermion bubbles in the bottom vertical rung of the ladder are not in the NLL anomalous dimension, but contribute to the scale of the coupling. See \cite{Thorne}.}, but the latter are missing in the exact kinematics calculation due the LL nature of the impact factor. However, one can see that the higher order $n_f$ terms do not constitute a very significant contribution relative to those at ${\cal O}(n_f)$.  
Similar expansions for the longitudinal coefficient are:
\begin{align}
\tilde{C}_{Lg}^{(2)DIS(e)}&=-\frac{5.333n_f}{N}+(-18.22n_f+.03292n_f^2)+(62.52n_f+.1427n_f^2)N\notag\\
&-(86.88n_f+.2551n_f^2)N^2+{\cal O}(N^3);\\
\tilde{C}_{Lg}^{(3)DIS(e)}&=\frac{409.5n_f}{N^2}+\frac{-1246n_f+1.727n_f^2}{N}+(2127n_f+40.14n_f^2+.01561n_f^3)\notag\\
&-(3436n_f+68.95n_f^2+.01892n_f^3)N+(5345n_f+89.73n_f^2\notag\\
&+.03326n_f^3)N^2+{\cal O}(N^3).
\end{align}
The complete results give:
\begin{align}
\tilde{C}_{Lg}^{(2)DIS}&=-\frac{5.333n_f}{N}+(-6.229n_f+.8889n_f^2)+(80.69n_f-4.850n_f^2)N\notag\\
&+(-133.8n_f+10.04n_f^2)N^2+{\cal O}(N^3);\\
\tilde{C}_{Lg}^{(3)DIS}&=\frac{409.5n_f}{N^2}+\frac{-2076n_f+102.4n_f^2}{N}+(4730n_f-340.8n_f^2-.1139fl_{11}^gn_f^2\notag\\
&+.5926n_f^3)+(-9211n_f+854.1n_f^2-.4340fl_{11}^gn_f^2-5.973n_f^3)N\notag\\
&+(20054n_f-2251n_f^2+.08264fl_{11}^gn_f^2+25.74n_f^3)N^2+{\cal O}(N^3),
\end{align}
where $fl_{11}^g=<e>^2/<e^2>$, taking averages over the active quark charges. The estimation of NLL terms is not as good as for $\gamma_{qg}$, even for the ${\cal O}(n_f)$ contribution. Again taking $n_f=4$, the NLL term in the NNLO coefficient is estimated to within $35\%$. Nevertheless, the exact kinematics results are in good qualitative agreement with the complete results. The term in $fl_{11}^g$ will not be estimated by the exact kinematics calculation due to missing diagrams of the type shown in figure \ref{fl11g}. Also, this term is not associated with a small $x$ divergence at ${\cal O}(\alpha_S^3)$. Higher order terms in $n_f$ and $fl_{11}^g$ are not very significant contributions. \\

\begin{figure}
\begin{center}
\scalebox{0.8}{\includegraphics{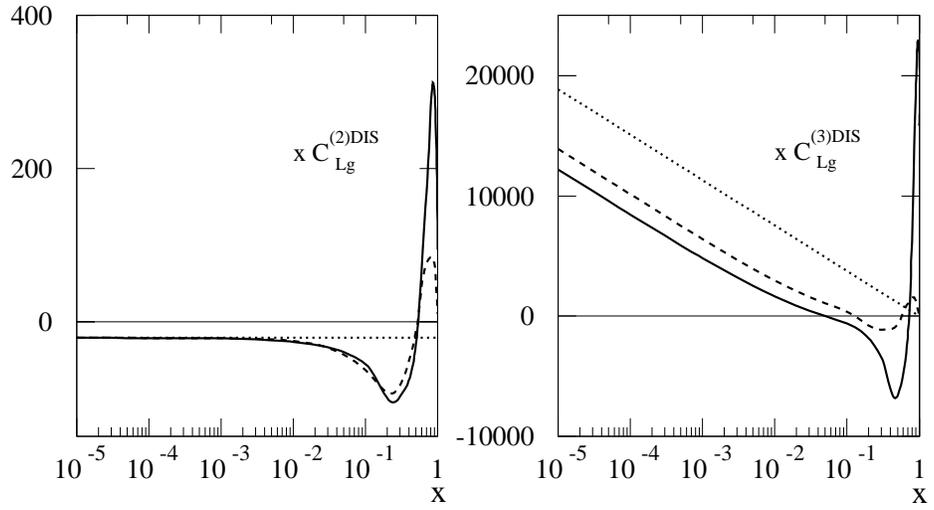}}
\caption{The DIS scheme NLO and NNLO longitudinal coefficient functions (solid) (multiplied by $x$), compared with the results from the LL impact photon-gluon impact factor with exact gluon kinematics (dashed). The LL approximation (dotted) is also shown.}
\label{clgplots}
\end{center}
\end{figure}

The $x$-space functions are shown in figure \ref{clgplots}.
Again the exact kinematics results have the same qualitative behaviour as the complete results at both small and large $x$, whereas the LL approximations are comparatively poor.\\

The greater accuracy in $\gamma_{qg}$ can in part be attributed to the derivative of $F_2$ in $\log(Q^2)$. In Mellin space, this amounts to multiplication of the sum of the transverse and longitudinal impact factors by $\gamma(N)$, which suppresses the differences noted above by $\alpha_S$.\\

\begin{figure}
\begin{center}
\scalebox{0.8}{\includegraphics{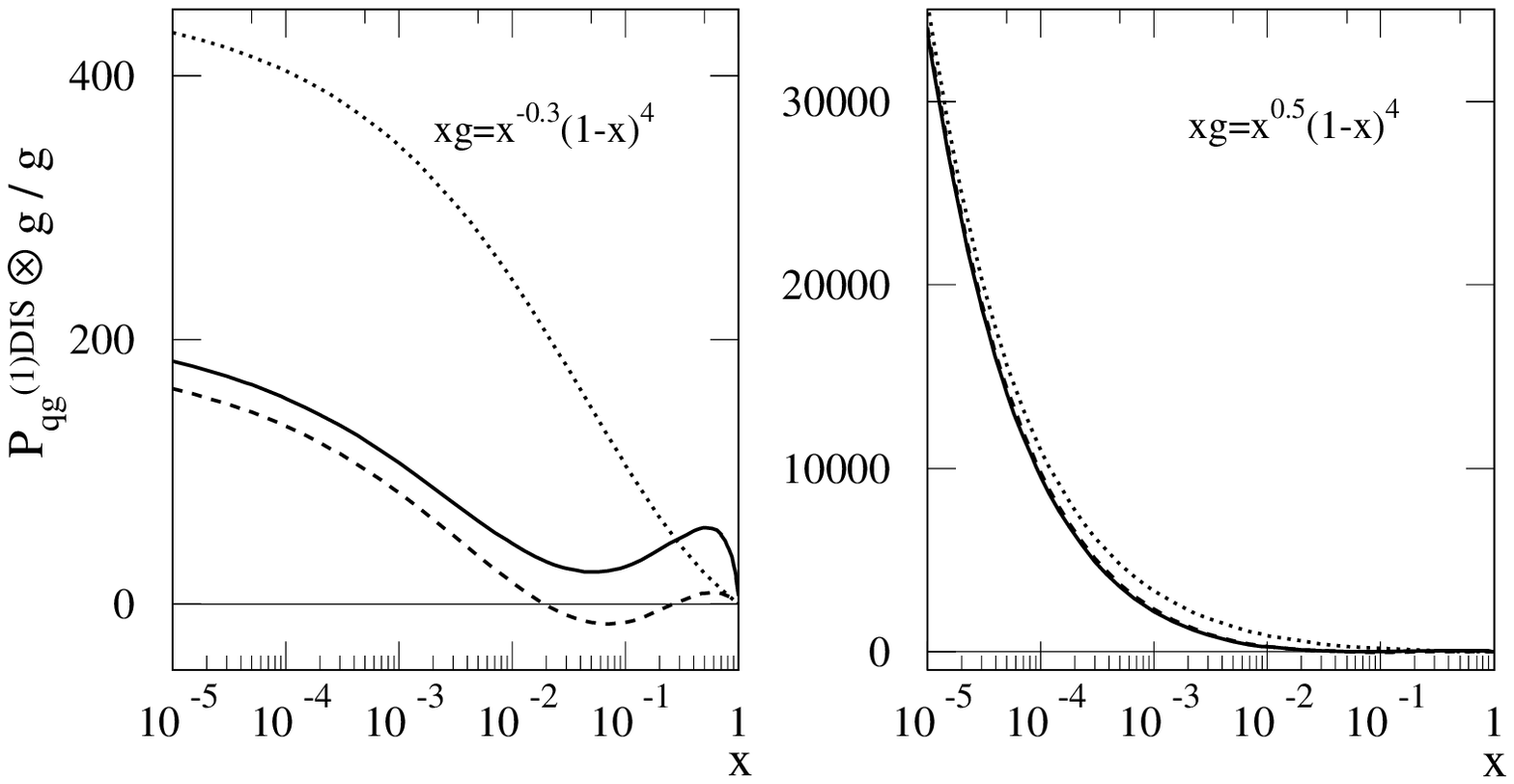}}
\caption{$(P_{qg}^{(1)DIS}(x)\otimes g(x,Q^2))/g(x,Q^2)$ with $n_f=4$, showing NLO (solid), exact kinematics (dashed) and LL (dotted) results.}
\label{ha2}
\end{center}
\end{figure}

\begin{figure}
\begin{center}
\scalebox{0.8}{\includegraphics{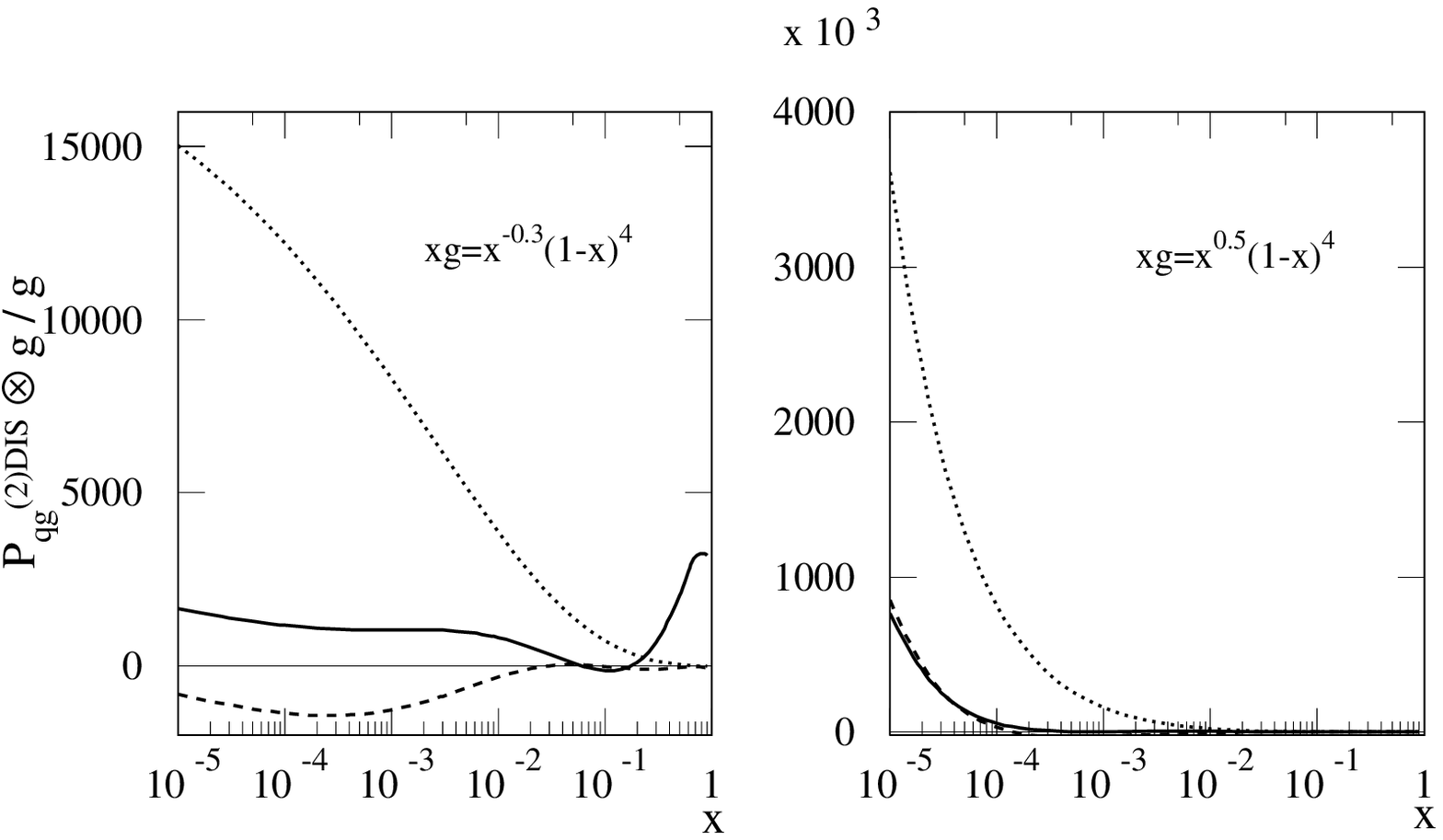}}
\caption{$(P_{qg}^{(2)DIS}(x)\otimes g(x,Q^2))/g(x,Q^2)$ with $n_f=4$, showing NNLO (solid), exact kinematics (dashed) and LL (dotted) results.}
\label{ha3}
\end{center}
\end{figure}

\begin{figure}
\begin{center}
\scalebox{0.8}{\includegraphics{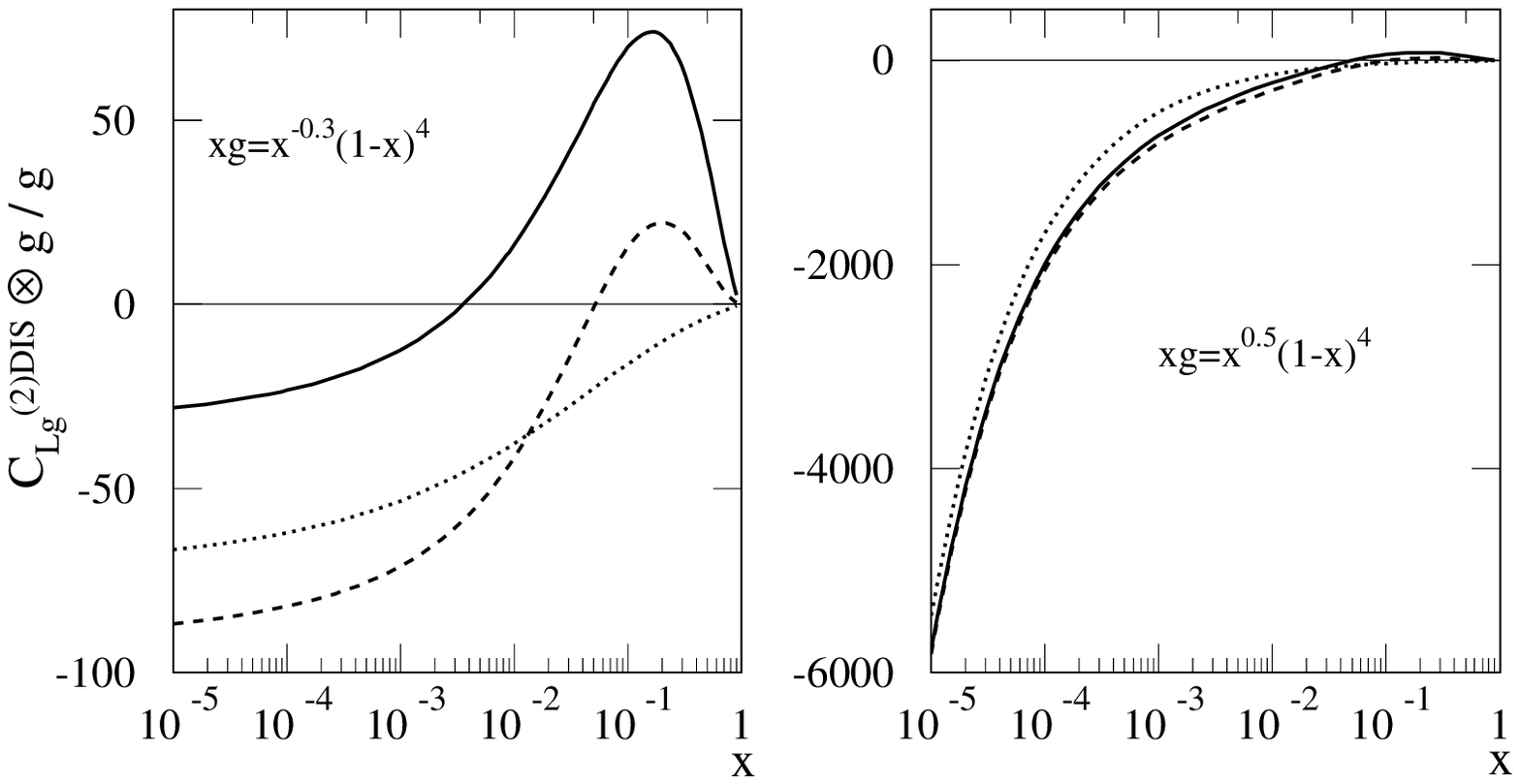}}
\caption{$C_{Lg}^{(2)DIS}\otimes g$ with $n_f=4$, showing NLO (solid), exact kinematics (dashed) and LL (dotted) results.}
\label{cl2}
\end{center}
\end{figure}

\begin{figure}
\begin{center}
\scalebox{0.8}{\includegraphics{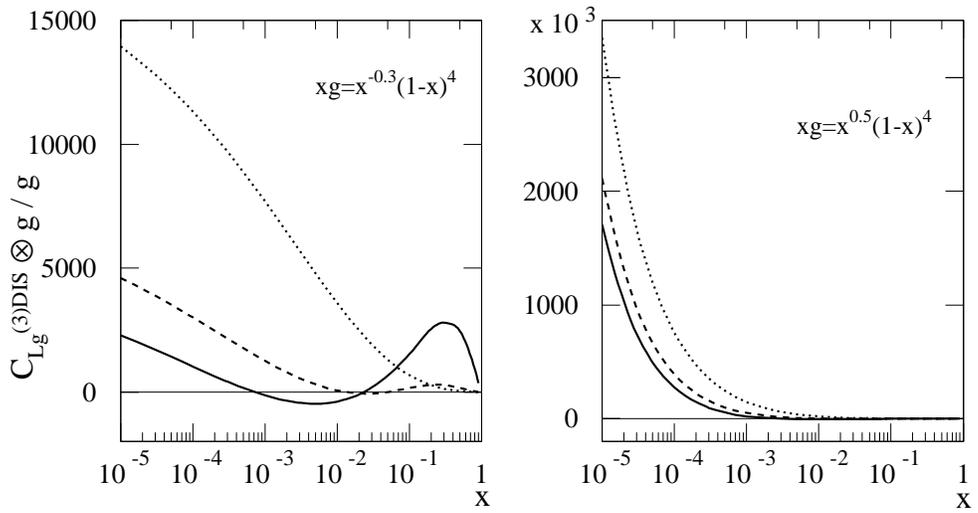}}
\caption{$C_{Lg}^{(3)DIS}\otimes g$ with $n_f=4$, showing NNLO (solid), exact kinematics (dashed) and LL (dotted) results.}
\label{cl3}
\end{center}
\end{figure}

The $x$-space functions will ultimately be convolved with parton distribution functions. Hence it is necessary to check the behaviour of the $x$-space exact kinematics expressions when convolved with a suitable gluon distribution. Following \cite{Vogtcl}, we convolve with the model gluon distributions:
\begin{align}
xg(x)&=x^{-0.3}(1-x)^4;\\
xg(x)&=x^{0.5}(1-x)^4,
\end{align}
where the former corresponds to a high $Q^2$ scale ($\simeq 30\text{GeV}^2$), and the latter reflects the fact that the gluon can be valence-like (or even negative at low $x$) at low $Q^2\simeq 1\text{GeV}^2$ \cite{MRST02,CTEQ}. One expects resummation of small $x$ terms to be more important at low $Q^2$, due to the higher value of $\alpha_S$. The results for $P_{qg}^{(1)}$ and $P_{qg}^{(2)}$ are shown in figures \ref{ha2} and \ref{ha3}.
Results for $C_{Lg}^{(2)DIS}\otimes g$ and $C_{Lg}^{(3)DIS}\otimes g$ are shown in figure \ref{cl2} and \ref{cl3}. 
For both the splitting and coefficient functions, the exact kinematics results qualitatively approximate the structure of the complete results. Comparing them with the LL terms at small $x$, after convolution with the gluon, one sees that they are much closer to the complete results. The exception is $C_{Lg}^{(2)}$, where the LL terms convolved with the gluon distribution are closer to the complete result at low $x$, aided by the fact that at this order the coefficient function has no next-to-leading small $x$ divergence. At NNLO, where NLL terms are present, the exact kinematics results perform better at small $x$. The exact kinematics and complete results generally agree more at the lower momentum scale. This is due to the less singular gluon distribution at low $Q^2$, and the small $x$ part of the coefficient playing a more dominant role. However, at higher $Q^2$, the effect of a more singular gluon distribution will be compensated in part by a lower value of $\alpha_S$, and resummation becomes less important.\\

The NNLO exact kinematics splitting function gives a negative result when convolved with the more singular gluon,  turning positive only at very low $x\simeq 10^{-7}$. This can be attributed to the large negative dip in the exact kinematics function (see figure \ref{GAMqg}) at intermediate $x$. Given that the gluon is more singular than the splitting function, the low $x$ limit of the convolution is dominated by both high and low $x$ information in the splitting function. To see how this works, consider the model splitting function:
\begin{equation}
P=\frac{A}{x}+B\delta(1-x),
\label{modelP}
\end{equation}
where the first and second terms give the dominant behaviour at small and high $x$ respectively. Consider convolving this with the following ``gluon'':
\begin{equation}
xf=x^{\alpha}\theta(x_0-x),
\label{modelg}
\end{equation}
which is singular or valence-like at low $x$ depending on whether $\alpha<0$ or $\alpha>0$, and vanishes at high $x$. One then has:
\begin{equation}
\frac{\partial\,(xf)}{\partial\log(Q^2)}=xP\otimes f=\left[B-\frac{A}{\alpha}\right]x^{\alpha}+\frac{Ax_0^{\alpha}}{\alpha}.
\label{model_convol}
\end{equation}
If $\alpha>0$, the small $x$ term in the splitting function dominates the convolution. If on the other hand $\alpha<0$, the bracketed term in equation (\ref{model_convol}) gives the leading small $x$ behaviour, which is a mixture of both the small and large $x$ terms of the splitting function. This also accounts for the lack of a common small $x$ limit in the left-hand plots of figures \ref{ha2}, \ref{ha3}, \ref{cl2}, \ref{cl3}, as each of the three splitting functions has a different high $x$ behaviour. Note that the complete NNLO $P_{qg}$ also has a negative dip at intermediate $x$. This leads to some negative behaviour after the convolution, but not for $x\lesssim 0.05$ in figure \ref{ha3}. 

\section{Conclusions}
We have shown that the imposition of exact gluon kinematics in the LL virtual photon-gluon impact factor gives a good approximation to the NLL parts of the NLO and NNLO splitting and coefficient functions $P_{qg}^{DIS}$ and $C_{Lg}^{DIS}$ up to ${\cal O}(\alpha_S^3)$. The qualitative behaviour is also good over the whole $x$ range. We see this both by examining poles in $N$-space and also convolving the $x$-space functions with suitable model gluon distributions. Hence in the absence of the full NLL impact factor \footnote{Calculation is, however, in progress \cite{Bartels1,Bartels2,Bartels3}.}, we have confidence that the exact kinematics results can be used for an accurate NLL analysis of the proton structure functions. \\

It may also be possible to impose exact kinematics in the impact factors for heavy quark production \cite{Catani2}. In this case, however, one needs to define a suitable factorisation scheme in order to interpret the impact factors in terms of splitting and coefficient functions.\\

The NNLO DIS scheme splitting and longitudinal coefficient functions (excluding the coefficients for charged current scattering) have been parameterised and presented here. There are significant qualitative differences with the $\msbar$ scheme results, particularly in the appearance of divergent high $x$ terms from soft gluon resummations in the splitting functions. These functions are available on request and can easily be applied for parton analyses in the DIS scheme at NNLO.

\section{Acknowledgements}
CDW would like to thank Jeppe Andersen for useful discussions, and is grateful to PPARC for a research studentship. RST thanks the Royal Society for the award of a University Research Fellowship.

\appendix
\section{Appendix: The exact kinematics splitting and coefficient functions}
The $N$-space splitting and coefficient functions derived from the exact kinematic impact factors involve the function $\psi(N)=\Gamma(N)/\Gamma'(N)$ and its derivatives, where $\Gamma(N)$ is the Euler gamma function. The $\psi$ functions can be expressed as analytically continued harmonic sums \cite{Blumlein}, which one can then inverse Mellin transform to $x$-space. For brevity we define:

\begin{eqnarray}
L_0=\log(x), & L_1=\log(1-x).
\end{eqnarray}
Then the results are:
\begin{align}
C_{Lg}^{(2)DIS(e)}(x)&=\left(-240x^2+272x)L_0-1196/3x^2-92-16/3x^{-1}+496x\right)n_f\notag\\
&+\left(32/27x(1-x)L_0-56/27x^2-8/27+64/27x^{-1}\right)n_f^2;
\end{align}
\begin{align}
C_{Lg}^{(3)DIS(e)}(x)&=\left([3468x-2700x^2]L_0^2+[-96x^{-1}L_1^2+(2312x-512x^{-1}-1800x^2)L_1-2820\right.\notag\\
&-11320x^2+13432x+(192\text{Li}_2(x)+32\pi^2-2176/3)x^{-1}]L_0+[900x^2+256x^{-1}\notag\\
&-1156x]L_1^2+[-1412+1800x+(32\pi^2-384-192\text{Li}_2(x))x^{-1}-4x^{-2}]L_1-1032\notag\\
&-40892/9x^2+6952x+[64\text{Li}_2(x)-192\text{Li}_3(1-x)-384\text{Li}_3(x)-12388/9\notag\\
&\left.+384\zeta(3)-32/3\pi^2]x^{-1}\right)n_f\notag\\
&+\left([272/9x-80/3x^2]L_0^2+[(-160/9x^2-64/27x^{-1}+544/27x)L_1-992/27x^2\right.\notag\\
&+64/9x-16]L_0+[-272/27x+32/27x^{-1}+80/9x^2]L_1^2+[160/9x-64/9x^{-1}\notag\\
&-304/27+16/27x^{-2}]L_1+1168/27+5024/81x^2-2800/27x+[-128/81\notag\\
&\left.+32/81\pi^2-64/27\text{Li}_2(x)]x^{-1}\right)n_f^2\notag\\
&+16/243\left(x(1-x)[3L_0^2+2L_1L_0+14L_0-L_1^2]-L_0-[x^{-2}+1-2x]L_1\right.\notag\\
&\left.-2-7x^2+10x-x^{-1}\right)n_f^3;
\end{align}
\begin{align}
P_{qg}^{(1)DIS(e)}(x)&=\left([92+120x^2-136x]L_0+1048/3x^2+44+104/3x^{-1}-384x\right)n_f\notag\\
&+\left([16/27x(x-1)^2+8/27]L_0+8/27+16/9x(x-1)\right)n_f^2;
\end{align}
\begin{align}
P_{qg}^{(2)DIS(e)}(x)&=\left([1350x^2-1734x+1587]L_0^2+[-96x^{-1}L_1^2+(-1156x+900x^2+1058-560x^{-1})L_1\right.\notag\\
&-890+10160x^2-11340x+(192\text{Li}_2(x)+32\pi^2-2272/3)x^{-1}]L_0+[578x-529\notag\\
&-450x^2+280x^{-1}]L_1^2+[706-900x+(32\pi^2-288-192\text{Li}_2(x))x^{-1}-2x^{-2}]L_1\notag\\
&+5992+86146/9x^2-12248x+[136/3\pi^2+384\zeta(3)-272\text{Li}_2(x)-29842/9-384\text{Li}_3(x)\notag\\
&\left.-192\text{Li}_3(1-x)]x^{-1}\right)n_f\notag\\
&+\left([40/3x^2+92/9-136/9x]L_0^2+[(184/27-64/27x^{-1}-272/27x+80/9x^2)L_1\right.\notag\\
&+1696/27x^2-1448/27-1184/27x]L_0+[-92/27+32/27x^{-1}+136/27x-40/9x^2]L_1^2\notag\\
&+[8/27x^{-2}+152/27-32/9x^{-1}-80/9x]L_1+800/27-472/81x^2+88/3x\notag\\
&\left.+[32/81\pi^2-4304/81-64/27\text{Li}_2(x)]x^{-1}\right)n_f^2\notag\\
&+4/729\left(2x(x-1)[3L_0^2+2L_0L_1-L_1^2]+3L_0^2+2L_1+2[3-22x+24x^2]L_0\right.\notag\\
&\left.-L_1^2+2[1-2x-x^{-2}]L_1+50x^2-56x+8-2x^{-1}\right)n_f^3,
\end{align}
where $\text{Li}_n(x)$ is the $n^\text{th}$ polylogarithm function.

\section{Appendix: The DIS scheme splitting and coefficient functions}
Here we present parameterisations of the coefficient and splitting functions at NNLO in the DIS scheme \cite{Altarelli}. For completeness, all singlet and non-singlet splitting functions are given. The longitudinal coefficient functions are given only for neutral current structure functions, as the $\overline{\text{MS}}$ scheme coefficient functions for charged current scattering have yet to be published. First we define:
\begin{eqnarray}
{\cal D}_n=\left[\frac{\log^n(1-x)}{1-x}\right]_+,&L_0=\log(x),&L_1=\log(1-x)
\end{eqnarray}
Then the results are:
\begin{align}
P_{ns}^{+(2)DIS}&\simeq 785.06{\cal D}_0-2974.4{\cal D}_1+645.33{\cal D}_2+14669.3758\delta(1-x)+1868.3+6601.3x+243.6x^2\notag\\
&-522.1x^3+77.391L_1^3+[-2771.56x+3059.61]L_1^2+[2695.85x+13750]L_1\notag\\
&+[1.5802-15.818x]L_0^4+83.639L_0^3+[83.48L_1+915.18]L_0^2+[-272.00x+2497.50\notag\\
&-750.9L_1+8314.3L_1^2+544.00/(1-x)]L_0\notag\\
&+n_f\left(-325.18{\cal D}_0+403.89{\cal D}_1-78.222{\cal D}_2-2150.5868\delta(1-x)+12.951+217.65x\right.\notag\\
&+358.28x^2+44.79x^3+.95867xL_0^4+[2.573x-5.6436]L_0^3+[-118.68+10.503L_1]L_0^2\notag\\
&+[-67.556/(1-x)-155.46L_1-503.89L_1^2+33.778x-327.76]L_0-4.6904L_1^3\notag\\
&\left.+[-152.43+167.97x]L_1^2+[-180.68x-1417.78]L_1\right)\notag\\
&+n_f^2\left(7.6750{\cal D}_0-11.457{\cal D}_1+2.3704{\cal D}_2+63.6358\delta(1-x)-4.8837-28.501x\right.\notag\\
&-17.293x^2-.24667xL_0^3+[1.1852x/(1-x)-.59259x+3.5556]L_0^2+[11.457\notag\\
&\left.+3.9506x/(1-x)-4.3457x+10.817L_1]L_0-2.0000L_1^2+33.160L_1\right);
\end{align}
\begin{align}
P_{ns}^{-(2)DIS}&\simeq 785.06{\cal D}_0-2974.4{\cal D}_1+645.33{\cal D}_2+14659\delta(1-x)-42.670+10704x+297.0x^2\notag\\
&-433.2x^3+1.4321L_0^4+106.84L_0^3+[994.40-860.64L_1]L_0^2+[-272.00x-630.82L_1\notag\\
&+2107.4+9310.9L_1^2+544.00/(1-x)]L_0+65.291L_1^3+[3085.1-2771.5x]L_1^2\notag\\
&+[14503.+2695.9x]L_1\notag\\
&+n_f\left(-325.18{\cal D}_0+403.89{\cal D}_1-78.222{\cal D}_2-2150.0\delta(1-x)+75.786+413.96x+77.89x^2\right.\notag\\
&+34.76x^3-[1.136x+7.4805]L_0^3+[.59212L_1-125.14]L_0^2+[-381.08L_1-67.556/(1-x)\notag\\
&+33.778x-564.29L_1^2-321.26]L_0-3.9570L_1^3+[167.97x-149.42]L_1^2\notag\\
&\left.-[1421.1+180.68x]L_1\right)\notag\\
&+n_f^2\left(7.6750{\cal D}_0-11.457{\cal D}_1+2.3704{\cal D}_2+63.585\delta(1-x)-2.0572-55.288x+[3.5846\notag\right.\\
&-.59259x+4.0479L_1+1.1852x/(1-x)]L_0^2+[23.959L_1+12.039+3.9506x/(1-x)\notag\\
&\left.-4.3457x]L_0-2.2760L_1^2+30.600L_1\right);
\end{align}
\begin{align}
P_{ps}^{(2)DIS}&\simeq n_f\left(-193299-672088x+104121x^2+964027x^3-201675x^4-1327.61x^{-1}-820.836L_0^3\right.\notag\\
&+[-61102.6L_1-9741.91]L_0^2+[307888L_1-190993L_1^2-75017.3-196.207x^{-1}]L_0+2332.86L_1\notag\\
&\left.-1876643L_1^2+10385.8L_1^3-1121.25xL_1+1876481xL_1^2-10509.5xL_1^3+24.88888L_1^4\right)\notag\\
&+n_f^2\left(530.035+7303.79x-937.737x^2-7925.90x^3+1105.83x^4+9.36593x^{-1}-1.65094L_0^4\right.\notag\\
&+1.51450L_0^3+[251.096L_1-31.8095]L_0^2+[-1502.08L_1+123.563+13313.6L_1^2]L_0+2649.16L_1\notag\\
&\left.+13574.4L_1^2-2639.02xL_1-13583.3xL_1^2+1.18519L_1^3\right);
\end{align}
\begin{align}
P_{gg}^{(2)DIS}&\simeq 2643.5{\cal D}_0+4425.8739\delta(1-x)-20852+3968x-3363x^2+4848x^3+14214x^{-1}-144L_0^4\notag\\
&+72L_0^3+[8757L_1-7471]L_0^2+[274.4+2675.8x^{-1}+7305L_1]L_0+3589L_1\notag\\
&+n_f\left(-412.172{\cal D}_0-534.1666\delta(1-x)+94680.9+423522x-62541.01x^2-569436x^3\right.\notag\\
&+120946x^4+1149.99x^{-1}+18.9631L_0^4+660.814L_0^3+[24297.9L_1+5133.55]L_0^2\notag\\
&+[1099250L_1^2-175012L_1+220.737x^{-1}+40461.3]L_0-24.8889L_1^4+[2062.11x-1913.21]L_1^3\notag\\
&\left.+[-1093524x+1093454]L_1^2+[-22404.4x+22442.5]L_1\right)\notag\\
&+n_f^2\left(-1.77778{\cal D}_0+6.44153\delta(1-x)-19903.1-81663.3x+11472.2x^2+114322x^3-24596.2x^4\right.\notag\\
&-17.8171x^{-1}-81.0657L_0^3+[-5570.25L_1-1006.77]L_0^2+[-7493.22-215788L_1^2+(85.25x\notag\\
&\left.+37019.4)L_1]L_0+[784.390-787.057x]L_1^3+[212770x-212747]L_1^2+[162.579-283.399x]L_1\right)\notag\\
&+n_f^3\left(14.399+15.108x-104.84x^2+41.797x^3+33.545x^4+.44376L_0^3+[-22.307L_1\right.\notag\\
&+2.6393]L_0^2+[-139.31L_1^2-112.57L_1+9.5276]L_0-.26473(x-1)L_1^3+140.68(x-1)L_1^2\notag\\
&\left.+112.93(x-1)L_1\right);
\end{align}
\begin{align}
P_{qg}^{(2)DIS}&\simeq n_f\left(5.2833\delta(1-x)-396354-1679228x+400583x^2+2086958x^3-413461x^4\right.\notag\\
&-3164.7x^{-1}+19.852L_0^4+[-1610.1-252.5x]L_0^3+[-130777L_1-18993]L_0^2\notag\\
&+[-152378+612746L_1-441.47x^{-1}-3977765L_1^2]L_0+36.004L_1^4+[2923.9-3143.9x]L_1^3\notag\\
&\left.+[4014417x-4013999]L_1^2+[325972x-323111]L_1\right)\notag\\
&+n_f^2\left(-0.2404\delta(1-x)-2697.2-16699x+11510x^2+8064.9x^3+30.195x^{-1}-10.237L_0^4\right.\notag\\
&+[-32.113+11.70x]L_0^3+[-1088.9L_1-427.67-98.07x]L_0^2+[2782.9L_1-1477.3\notag\\
&\left.-28176L_1^2]L_0+2.6670L_1^3+[29629x-29652]L_1^2+[7440.6x-7360.6]L_1\right)\notag\\
&+n_f^3\left(0.0013\delta(1-x)+143.623-835.290x+540.973x^2+150.487x^3+.326927L_0^4+3.75075L_0^3\right.\notag\\
&+[21.1805L_1+25.3458]L_0^2+[95.2390+196.232L_1-865.734L_1^2]L_0+881.848(x-1)L_1^2\notag\\
&\left.+32.8575(x-1)L_1\right);
\end{align}
\begin{align}
P_{gq}^{(2)DIS}&\simeq-10172.599{\cal D}_0+2619.956{\cal D}_1+3026.479{\cal D}_2-75.85220{\cal D}_3-118.5187{\cal D}_4\notag\\
&-17666.5673\delta(1-x)-63856.3-226976x+5645.05x^2+370971x^3-80184.6x^4+6133.90x^{-1}\notag\\
&-52.9383L_0^4-269.675L_0^3+[-972.9x-16883.4L_1-6887.42]L_0^2+[-25459.0+1189.3x^{-1}\notag\\
&-692833L_1^2+122913L_1]L_0+89.4494L_1^4+[7443.02-7918.33x]L_1^3+[-666832+658887x]L_1^2\notag\\
&+[16626.4-32060.8x]L_1\notag\\
&+n_f\left(935.1848{\cal D}_0-550.4791{\cal D}_1-56.29637{\cal D}_2+25.28401{\cal D}_3+2589.9531\delta(1-x)+35445+73884x\right.\notag\\
&-11203x^2-127979x^3+27317x^4+350.55x^{-1}+4.7407L_0^4+312.26L_0^3+[9521.1L_1+108.6x\notag\\
&+2357.5]L_0^2+[99.282x^{-1}-45381L_1+16599+260063L_1^2]L_0-14.222L_1^4+[-1762.4\notag\\
&\left.+1847.8x]L_1^3+[-254267x+254637]L_1^2+[-4256.9+4324.8x]L_1\right)\notag\\
&+n_f^2\left(-12.698{\cal D}_0+17.185{\cal D}_1-3.5556{\cal D}_2-93.6748\delta(1-x)-103.10+809.52x-655.80x^2\right.\notag\\
&-5.0491x^{-1}-8.0350L_0^3+[31.758L_1-20.430]L_0^2+[-273.77L_1+317.30L_1^2-144.58]L_0\notag\\
&\left.-3.5556L_1^3+[-230.63x+241.15+3.5556x^{-1}]L_1^2+[96.502x+11.852x^{-1}-110.71]L_1\right);
\end{align}
\begin{align}
C_{Lns}^{(3)+DIS}(x)&\simeq-3634.5+5025.2x-614.77x^3-996.21x^4+(1-x)[8452.3L_1+4090.2L_1^2+175.59L_1^3\notag\\
&+225.30L_1^4]-3280.3L_0L_1-1082.7L_0^2L_1-911.45L_0-81.823L_0^2-.72047L_0^3-1780.0L_0L_1^2\notag\\
&-4059.2L_1+125.02L_1^2+21.113L_1^3+1.6059L_1^4\notag\\
&+n_f\left(617.05-1670.5x+(1-x)[23.584L_1-106.82L_1^2]+1717.8L_0L_1\right.\notag\\
&\left.+465.96L_0^2L_1+171.90L_0+6.9942L_0^2+370.25L_1-45.190L_1^2\right)\notag\\
&+n_f^2\left(-17.038+35.968x+(1-x)[-22.215L_1+23.829L_1^2]-66.179L_0L_1-12.884L_0^2L_1\right.\notag\\
&\left.-5.1888L_0-.025315L_0^2+24.794L_0L_1^2-15.012L_1+2.3704L_1^2\right)\notag\\
&+fl_{11}^{ns}n_f\left([107.0+321.05x-54.62x^2](1-x)-26.717+9.773L_0\right.\notag\\
&\left.+[363.8+68.32L_0]xL_0-320/81L_0^2[2+L_0]\right)x;
\end{align}
\begin{align}
C_{Lps}^{(3)DIS}(x)&\simeq n_f\left(1769.7-441.62x-[182.00L_0+899.64]x^{-1}-(1-x)[23.584L_1-106.82L_1^2]\right.\notag\\
&+53648L_0L_1+11604L_0^2L_1-894.81L_0+105.36L_0^2+(1-x)[81652L_1+3880.7L_1^2]\notag\\
&\left.-76.310L_0^3-8700.9L_0L_1^2\right)\notag\\
&+n_f^2\left(4087.2-4143.1x+47.29x^{-1}+(1-x)[2293.9L_1+654.38L_1^2]+978.21L_0L_1\right.\notag\\
&+2199.0L_0^2L_1+1484.3L_0+176.52L_0^2+18.327L_0^3+511.80L_0L_1^2\notag\\
&\left.-(1-x)[-22.215L_1+23.829L_1^2]\right)\notag\\
&+fl_{11}^{ps}n_f\left([107.0+321.05x-54.62x^2](1-x)-26.717+9.773L_0\right.\notag\\
&\left.+[363.8+68.32L_0]xL_0-320/81L_0^2[2+L_0]\right)x;
\end{align}
\begin{align}
C_{Lg}^{(3)DIS}(x)&\simeq n_f\left(-4573.1+77228x-70637x^3-[409.506L_0+2076.4]x^{-1}\right.\notag\\
&+(1-x)[-8666.9L_1+267612L_1^2-4500.1L_1^3]-8146.1L_0L_1+4257.5L_0^2L_1\notag\\
&\left.-4277.2L_0-241.08L_0^2-246.51L_0^3+272818L_0L_1^2+.32800L_1\right)\notag\\
&+n_f^2\left(8878.1-14399x+5430.1x^3+102.40x^{-1}+(1-x)[5143.2L_1-83.489L_1^2]\right.\notag\\
&\left.+7051.6L_0L_1+5593.0L_0^2L_1+3258.9L_0+481.19L_0^2+68.034L_0^3+516.40L_0L_1^2\right)\notag\\
&+n_f^3\left(-287.66-494.46x+782.72x^3+(1-x)[-614.31L_1-1547.3L_1^2]-32.680L_0L_1\right.\notag\\
&\left.-112.24L_0^2L_1-132.42L_0-26.899L_0^2-2.6004L_0^3-1490.4L_0L_1^2\right)\notag\\
&+fl_{11}^gn_f^2\left([-0.0105L_1^3+1.550L_1^2+19.72xL_1-66.745x+0.615x^2](1-x)\right.\notag\\
&+20/27xL_0^4+[280/81+2.260x]xL_0^3-[15.40-2.201x]xL_0^2\notag\\
&\left.-[71.66-0.121x]xL_0\right).
\end{align}

\bibliography{refs}
\end{document}